\documentstyle[fleqn,11pt]{book}
\begin{document}
\parindent 1.4cm
\large
\vspace{0.2cm}
\begin{center}
{{\bf The Quantum Wave Packet of the Schr\"{o}dinger's Equation for
Continuous Quantum Measurements}}
\end{center}
\begin{center}
{{\bf J. M. F. Bassalo$^{1}$,\ P. T. S. Alencar$^{2}$,\  D. G. da
Silva$^{3}$,\ A. Nassar$^{4}$\ and\ M. Cattani$^{5}$}}
\end{center}
\begin{center}
{$^{1}$\ Funda\c{c}\~ao Minerva,\ R. Serzedelo Correa 347, 1601\
- CEP\ 66035-400,\ Bel\'em,\ Par\'a,\ Brasil}
\end{center}
\begin{center}
{E-mail:\ bassalo@amazon.com.br}
\end{center}
\begin{center}
{$^{2}$\ Universidade Federal do Par\'a\ -\ CEP\ 66075-900,\ Guam\'a,
Bel\'em,\ Par\'a,\ Brasil}
\end{center}
\begin{center}
{E-mail:\ tarso@ufpa.br}
\end{center}
\begin{center}
{$^{3}$\ Escola Munguba do Jari, Vit\'oria do Jari\ -\ CEP\
68924-000,\ Amap\'a,\ Brasil}
\end{center}
\begin{center}
{E-mail:\ danielgemaque@yahoo.com.br}
\end{center}
\begin{center}
{$^{4}$\ Extension Program-Department of Sciences, University of California,\
Los Angeles, California 90024,\ USA}
\end{center}
\begin{center}
{E-mail:\ nassar@ucla.edu}
\end{center}
\begin{center}
{$^{5}$\ Instituto de F\'{\i}sica da Universidade de S\~ao Paulo. C. P.
66318, CEP\ 05315-970,\ S\~ao Paulo,\ SP, Brasil}
\end{center}
\begin{center}
{E-mail:\ mcattani@if.usp.br}
\end{center}
\par
{\bf Abstract}:\ In this paper we study the quantum wave packet of the
Schr\"{o}dinger's equation for continuous quantum measurements.
\vspace{0.2cm}
\vspace{0.2cm}
\par
{\bf 1.\ Introduction}
\vspace{0.2cm}
\par
In this paper we will study the wave packet of a Schr\"{o}dinger
Continuous Measurements Equation proposed by Nassar [1], using the
quantum mechanical formalism of de Broglie-Bohm.[2]
\vspace{0.2cm}
\par
{\bf 2.\ The Continuous Measurement Schr\"{o}dinger's Equation}
\vspace{0.2cm}
\par
According to Nassar [1] a Schr\"{o}dinger equation assuming continuous
measurements is given by:
\begin{center}
{i\ ${\hbar}\ {\frac {{\partial}{\Psi}(x,\ t)}{{\partial}t}}\ =\ -\
{\frac {{\hbar}^{2}}{2\ m}}\ {\frac {{\partial}^{2}\ {\Psi}(x,\
t)}{{\partial}x^{2}}}\ +\ {\Big {[}}\ {\frac {1}{2}}\ m\ {\Omega}^{2}(t)\
x^{2}\ +\ {\lambda}\ x\ X(t)\ {\Big {]}}\ {\Psi}(x,\ t)\ -$}
\end{center}
\begin{center}
{$-\ {\frac {i\ {\hbar}}{4\ {\tau}}}\ {\Bigg {(}}\ {\frac {[x\ -\
q(t)]^{2}}{{\delta}^{2}(t)}}\ -\ 1\ {\Bigg {)}}\ {\Psi}(x,\ t)$\ ,\ \ \
\ \ (2.1)}
\end{center}
where ${\Psi}(x,\ t)$ is a wave function which describes a given
system, $X(t)$ is the position of a classical particle submitted to a
time dependent harmonic potential with frequency ${\Omega(t)}$, and
${\tau}$ and ${\delta}$ have dimensions of space and time,
respectively, and $q(t)$ is average value $<\ x(t)\ >$.
\par
Writing the wave function ${\Psi}(x,\ t)$ in the polar form defined by
the Madelung-Bohm transformation [3,4] we obtain:
\begin{center}
{${\Psi}(x,\ t)\ =\ {\phi}(x,\ t)\ e^{i\ S(x,\ t)}$\ ,\ \ \ \ \ (2.2)}
\end{center}
where $S(x\ ,t)$ is the classical action and ${\phi}(x,\ t)$ will be
defined in what follows.
\par
Calculating the derivatives, temporal and spatial, of (2.2), we get:
\begin{center}
{${\frac {{\partial}{\Psi}}{{\partial}t}}\ =\ e^{i\ S}\ {\frac
{{\partial}{\phi}}{{\partial}t}}\ +\ i\ {\phi}\ e^{i\ S}\ {\frac
{{\partial}S}{{\partial}t}}\ \ \ {\to}$}
\end{center}
\begin{center}
{${\frac {{\partial}{\Psi}}{{\partial}t}}\ =\ e^{i\ S}\ ({\frac
{{\partial}{\phi}}{{\partial}t}}\ +\ i\ {\phi}\ {\frac
{{\partial}S}{{\partial}t}})\ =\ (i\ {\frac
{{\partial}S}{{\partial}t}}\ +\ {\frac {1}{{\phi}}}\ {\frac
{{\partial}{\phi}}{{\partial}t}})\ {\Psi}$\ ,\ \ \ \ \ (2.3a,b)}
\end{center}
\begin{center}
{${\frac {{\partial}{\Psi}}{{\partial}x}}\ =\ e^{i\ S}\ ({\frac
{{\partial}{\phi}}{{\partial}x}}\ +\ i\ {\phi}\ {\frac
{{\partial}S}{{\partial}x}})\ =\ (i\ {\frac
{{\partial}S}{{\partial}x}}\ +\ {\frac {1}{{\phi}}}\ {\frac
{{\partial}{\phi}}{{\partial}x}})\ {\Psi}$\ ,\ \ \ \ \ (2.3c,d)}
\end{center}
\begin{center}
{${\frac {{\partial}^{2}{\Psi}}{{\partial}x^{2}}}\ =\ {\frac
{{\partial}}{{\partial}x}}\ [(i\ {\frac {{\partial}S}{{\partial}x}}\ +\
{\frac {1}{{\phi}}}\ {\frac {{\partial}{\phi}}{{\partial}x}})\
{\Psi}]$\ =}
\end{center}
\begin{center}
{=\ ${\Psi}\ [i\ {\frac {{\partial}^{2}S}{{\partial}x^{2}}}\ +\ {\frac
{1}{{\phi}}}\ {\frac {{\partial}^{2}{\phi}}{{\partial}x^{2}}}\ -\
{\frac {1}{{\phi}^{2}}}\ ({\frac
{{\partial}{\phi}}{{\partial}x}})^{2}]\ +\ (i\ {\frac
{{\partial}S}{{\partial}x}}\ +\ {\frac {1}{{\phi}}}\ {\frac
{{\partial}{\phi}}{{\partial}x}})\ {\frac
{{\partial}{\Psi}}{{\partial}x}}$\ =}
\end{center}
\begin{center}
{=\ ${\Psi}\ [i\ {\frac {{\partial}^{2}S}{{\partial}x^{2}}}\ +\ {\frac
{1}{{\phi}}}\ {\frac {{\partial}^{2}{\phi}}{{\partial}x^{2}}}\ -\
{\frac {1}{{\phi}^{2}}}\ ({\frac
{{\partial}{\phi}}{{\partial}x}})^{2}]\ +\ (i\ {\frac
{{\partial}S}{{\partial}x}}\ +\ {\frac {1}{{\phi}}}\ {\frac
{{\partial}{\phi}}{{\partial}x}})\ (i\ {\frac
{{\partial}S}{{\partial}x}}\ +\ {\frac {1}{{\phi}}}\ {\frac
{{\partial}{\phi}}{{\partial}x}})\ {\Psi}$\ =}
\end{center}
\begin{center}
{=\ ${\Psi}\ [i\ {\frac {{\partial}^{2}S}{{\partial}x^{2}}}\ +\ {\frac
{1}{{\phi}}}\ {\frac {{\partial}^{2}{\phi}}{{\partial}x^{2}}}\ -\
{\frac {1}{{\phi}^{2}}}\ ({\frac {{\partial}{\phi}}{{\partial}x}})^{2}\
-\ ({\frac {{\partial}S}{{\partial}x}})^{2}\ +\ {\frac
{1}{{\phi}^{2}}}\ ({\frac {{\partial}{\phi}}{{\partial}x}})^{2}\ +\ 2\
{\frac {i}{{\phi}}}\ {\frac {{\partial}S}{{\partial}x}}\ {\frac
{{\partial}{\phi}}{{\partial}x}}]\ \ \ {\to}$}
\end{center}
\begin{center}
{${\frac {{\partial}^{2}{\Psi}}{{\partial}x^{2}}}\ =\ e^{i\ S}\ [{\frac
{{\partial}^{2}{\phi}}{{\partial}x^{2}}}\ +\ 2\ i\ {\frac
{{\partial}S}{{\partial}x}}\ {\frac {{\partial}{\phi}}{{\partial}x}}\
+\ i\ {\phi}\ {\frac {{\partial}^{2}S}{{\partial}x^{2}}}\ -\ {\phi}\
({\frac {{\partial}S}{{\partial}x}})^{2}]\ =$}
\end{center}
\begin{center}
{$=\ {\big {[}}\ i\ {\frac {{\partial}^{2}S}{{\partial}x^{2}}}\ +\ {\frac
{1}{{\phi}}}\ {\frac {{\partial}^{2}{\phi}}{{\partial}x^{2}}}\ -\
({\frac {{\partial}S}{{\partial}x}})^{2}\ +\ 2\ i\ {\frac {1}{{\phi}}}\
{\frac {{\partial}S}{{\partial}x}}\ {\frac
{{\partial}{\phi}}{{\partial}x}}\ {\big {]}}\ {\Psi}$\ .\ \ \ \ \ (2.3e,f)}
\end{center}
\par
Now, inserting the relations defined by eq. (2.3a,e) into eq. (2.1) we
have, remembering that $e^{i\ S}$ is common factor:
\begin{center}
{$i\ {\hbar}\ ({\frac {{\partial}{\phi}}{{\partial}t}}\ +\ i\ {\phi}\
{\frac {{\partial}S}{{\partial}t}})\ =\ -\ {\frac {{\hbar}^{2}}{2\ m}}\
[{\frac {{\partial}^{2}{\phi}}{{\partial}x^{2}}}\ +$}
\end{center}
\begin{center}
{$\ +\ 2\ i\ {\frac {{\partial}S}{{\partial}x}}\ {\frac
{{\partial}{\phi}}{{\partial}x}}\ +\ i\ {\phi}\ {\frac
{{\partial}^{2}S}{{\partial}x^{2}}}\ -\ {\phi}\ ({\frac
{{\partial}S}{{\partial}x}})^{2}]\ +\ {\big {[}}\ {\frac
{1}{2}}\ m\ {\Omega}(t)^{2}\ x^{2}\ +\ {\lambda}\ x\ X(t)\ {\big {]}}\
{\phi}\ -$}
\end{center}
\begin{center}
{$-\ {\frac {i\ {\hbar}}{4\ {\tau}}}\ {\big {(}}\ {\frac {[x\ -\
q(t)]^{2}}{{\delta}^{2}(t)}}\ -\ 1\ {\big {)}}\ {\phi}$\ ,\ \ \
\ \ (2.4)}
\end{center}
\par
Separating the real and imaginary parts of the relation (2.4), results:
\par
a)\ {\underline {imaginary part}}
\begin{center}
{${\frac {{\partial}{\phi}}{{\partial}t}}\ =\ -\ {\frac {{\hbar}}{2\
m}}\ {\big {(}}\ 2\ {\frac {{\partial}S}{{\partial}x}}\ {\frac
{{\partial}{\phi}}{{\partial}x}}\ +\ {\phi}\ {\frac
{{\partial}^{2}S}{{\partial}x^{2}}}\ {\big {)}}\ -\ {\frac {1}{4\
{\tau}}}\ {\big {(}}\ {\frac {[x\ -\ q(t)]^{2}}{{\delta}^{2}(t)}}\ -\
1\ {\big {)}}\ {\phi}$\ ,\ \ \ \ \ (2.5)}
\end{center}
\par
b)\ {\underline {real part}
\begin{center}
{-\ ${\hbar}\ {\phi}\ {\frac {{\partial}S}{{\partial}t}}\ =\ -\ {\frac
{{\hbar}^{2}}{2\ m}}\ [{\frac {{\partial}^{2}{\phi}}{{\partial}x^{2}}}\
-\ {\phi}\ ({\frac {{\partial}S}{{\partial}x}})^{2}]\ +\ {\big {[}}\ {\frac
{1}{2}}\ m\ {\Omega}(t)^{2}\ x^{2}\ +\ {\lambda}\ x\ X(t)\ {\big {]}}\
{\phi}\ \ \ ({\div}\ m\ {\phi})\ \ \ {\to}$}
\end{center}
\begin{center}
{-\ ${\frac {{\hbar}}{m}}\ {\frac {{\partial}S}{{\partial}t}}\ =\ -\ {\frac
{{\hbar}^{2}}{2\ m^{2}}}\ {\frac {1}{{\phi}}}\ {\big {[}}\ {\frac
{{\partial}^{2}{\phi}}{{\partial}x^{2}}}\ -\ {\phi}\ ({\frac
{{\partial}S}{{\partial}x}})^{2}\ {\big {]}}\ +\ {\big {[}}\ {\frac
{1}{2}}\ {\Omega}(t)^{2}\ x^{2}\ +\ {\frac {{\lambda}}{m}}\ x\ X(t)\
{\big {]}}$\ .\ \ \ \ \ (2.6)}
\end{center}
\vspace{0.2cm}
\par
{\bf Dynamics of the Schr\"{o}dinger's Equation for
Continuous Quantum Measurements}
\vspace{0.2cm}
\par
Now, let us see the correlation between the expressions (2.5-6) and the
traditional equations of the Ideal Fluid Dynamics [5]\ a) continuity
equation, b) Euler's equation. To do this let us perform the following
correspondences:
\begin{center}
{{\underline {Quantum density probability}}:\ \ \ \ \ ${\mid}\
{\Psi}(x,\ t)\ {\mid}^{2}\ =\ {\Psi}^{*}(x,\ t)\ {\Psi}(x,\ t)\ \ \ \ \
\ {\longleftrightarrow}$}
\end{center}
\begin{center}
{{\underline {Quantum mass density}}:\ \ \ \ \ ${\rho}(x,\ t)\ =\
{\phi}^{2}(x,\ t)\ \ \ {\longleftrightarrow}\ \ \ {\sqrt {{\rho}}}\ =\
{\phi}$\ ,\ \ \ \ \ (2.7a,b)}
\end{center}
\begin{center}
{{\underline {Gradient of the wave function}}:\ \ \ \ \
${\frac {{\hbar}}{m}}\ {\frac {{\partial}S(x,\ t)}{{\partial}x}}\ \ \ \
\ {\longleftrightarrow}$}
\end{center}
\begin{center}
{{\underline {Quantum velocity}}:\ \ \ \ \ $v_{qu}(x,\ t)\
{\equiv}\ v_{qu}$\ .\ \ \ \ \ (2.8)}
\end{center}
\par
Putting the relations (2.7b, 2.8) into the equation (2.5) we get:
\begin{center}
{${\frac {{\partial}{\sqrt {{\rho}}}}{{\partial}t}}\ =\ -\ {\frac
{{\hbar}}{2\ m}}\ {\big {(}}\ 2\ {\frac {{\partial}S}{{\partial}x}}\
{\frac {{\partial}{\sqrt {{\rho}}}}{{\partial}x}}\ +\ {\sqrt {{\rho}}}\
{\frac {{\partial}^{2}S}{{\partial}x^{2}}}\ {\big {)}}\ -\ {\frac
{1}{4\ {\tau}}}\ {\big {(}}\ {\frac {[x\ -\
q(t)]^{2}}{{\delta}^{2}(t)}}\ -\ 1\ {\big {)}}\ {\sqrt {{\rho}}}\ \ \
{\to}$}
\end{center}
\begin{center}
{${\frac {1}{2\ {\sqrt {{\rho}}}}}\ {\frac
{{\partial}{\rho}}{{\partial}t}}\ =\ -\ {\frac {{\hbar}}{2\ m}}\ {\big
{(}}\ 2\ {\frac {{\partial}S}{{\partial}x}}\ {\frac {1}{2\ {\sqrt
{{\rho}}}}}\ {\frac {{\partial}{\rho}}{{\partial}x}}\ +\ {\sqrt
{{\rho}}}\ {\frac {{\partial}^{2}S}{{\partial}x^{2}}}\ {\big {)}}\ -\
{\frac {1}{4\ {\tau}}}\ {\big {(}}\ {\frac {[x\ -\
q(t)]^{2}}{{\delta}^{2}(t)}}\ -\ 1\ {\big {)}}\ {\sqrt {{\rho}}}\ \ \
{\to}$}
\end{center}
\begin{center}
{${\frac {1}{{\rho}}}\ {\frac
{{\partial}{\rho}}{{\partial}t}}\ =\ -\ {\frac {{\hbar}}{m}}\ {\big
{(}}\ {\frac {{\partial}S}{{\partial}x}}\ {\frac {1}{{\rho}}}\
{\frac {{\partial}{\rho}}{{\partial}x}}\ +\ {\frac
{{\partial}^{2}S}{{\partial}x^{2}}}\ {\big {)}}\ -\ {\frac {1}{2\
{\tau}}}\ {\big {(}}\ {\frac {[x\ -\ q(t)]^{2}}{{\delta}^{2}(t)}}\ -\
1\ {\big {)}}\ \ \ {\to}$}
\end{center}
\begin{center}
{${\frac {1}{{\rho}}}\ {\frac
{{\partial}{\rho}}{{\partial}t}}\ =\ -\ {\frac
{{\partial}}{{\partial}x}}\ {\big {(}}\ {\frac {{\hbar}}{m}}\ {\frac
{{\partial}S}{{\partial}x}}\ {\big {)}}\ -\ {\frac {1}{{\rho}}}\ {\big
{(}}\ {\frac {{\hbar}}{m}}\ {\frac {{\partial}S}{{\partial}x}}\ {\big
{)}}\ {\frac {{\partial}{\rho}}{{\partial}x}}\ -\ {\frac {1}{2\
{\tau}}}\ {\big {(}}\ {\frac {[x\ -\ q(t)]^{2}}{{\delta}^{2}(t)}}\ -\
1\ {\big {)}}\ \ \ {\to}$}
\end{center}
\begin{center}
{${\frac {1}{{\rho}}}\ {\frac
{{\partial}{\rho}}{{\partial}t}}\ =\ -\ {\frac
{{\partial}v_{qu}}{{\partial}x}}\ -\ {\frac {v_{qu}}{{\rho}}}\ {\frac
{{\partial}{\rho}}{{\partial}x}}\ -\ {\frac {1}{2\ {\tau}}}\ {\big
{(}}\ {\frac {[x\ -\ q(t)]^{2}}{{\delta}^{2}(t)}}\ -\ 1\ {\big {)}}\ \
\ {\to}$}
\end{center}
\begin{center}
{${\frac {{\partial}{\rho}}{{\partial}t}}\ +\ {\rho}\ {\frac
{{\partial}v_{qu}}{{\partial}x}}\ +\ v_{qu}\ {\frac
{{\partial}{\rho}}{{\partial}x}}\ =\ -\ {\frac {{\rho}}{2\ {\tau}}}\ {\big
{(}}\ {\frac {[x\ -\ q(t)]^{2}}{{\delta}^{2}(t)}}\ -\ 1\ {\big {)}}\ \
\ {\to}$}
\end{center}
\begin{center}
{${\frac {{\partial}{\rho}}{{\partial}t}}\ +\ {\frac
{{\partial}({\rho}\ v_{qu})}{{\partial}x}}\ =\ -\ {\frac {{\rho}}{2\
{\tau}}}\ {\big {(}}\ {\frac {[x\ -\ q(t)]^{2}}{{\delta}^{2}(t)}}\ -\
1\ {\big {)}}$\ ,\ \ \ \ \ (2.9)}
\end{center}
which represents the continuity equation of the mass conservation law
of the Fluid Dynamics. We must note that this expression also
indicates descoerence of the considered physical system represented by
the eq. (2.1).\ Using eq. (2.7b) we now define the quantum potential
$V_{qu}$:
\begin{center}
{$V_{qu}(x,\ t)\ {\equiv}\ V_{qu}\ =\ -\ ({\frac {{\hbar}^{2}}{2\ m\
{\phi}}})\ {\frac {{\partial}^{2}{\phi}}{{\partial}x^{2}}}\ =\ -\
{\frac {{\hbar}^{2}}{2\ m}}\ {\frac {1}{{\sqrt {{\rho}}}}}\ {\frac
{{\partial}^{2}{\sqrt {{\rho}}}}{{\partial}x^{2}}}$\ ,\ \ \ \ \ (2.10a,b)}
\end{center}
the expression (2.6) will written as:
\begin{center}
{${\frac {{\hbar}}{m}}\ {\frac {{\partial}S}{{\partial}t}}\ +\ {\frac
{{\hbar}^{2}}{2\ m^{2}}}\ ({\frac {{\partial}S}{{\partial}x}})^{2}\ =\
-\ {\frac {1}{m}}\ {\Big {(}}\ {\big {[}}\ {\frac
{1}{2}}\ m\ {\Omega}^{2}(t)\ x^{2}\ +\ {\lambda}\ x\ X(t)\
{\big {]}}\ +\ V_{qu}\ {\Big {)}}$\ ,\ \ \ \ \ (2.11)}
\end{center}
or, using (2.8):
\begin{center}
{${\hbar}\ {\frac {{\partial}S}{{\partial}t}}\ +\ {\frac {1}{2}}\ m\
v_{qu}^{2}\ +\ {\frac {1}{2}}\ m\ {\Omega}^{2}(t)\ x^{2}\
+\ {\lambda}\ x\ X(t)\ +\ V_{qu}\ =$}
\end{center}
\begin{center}
{$=\ {\hbar}\ {\frac {{\partial}S}{{\partial}t}}\ +\ {\frac {1}{2}}\ m\
v_{qu}^{2}\ +\ V\ +\ V_{qu}\ =\ 0$\ .\ \ \ \ \ (2.12a,b)}
\end{center}
\par
Differentiating the relation (2.12a) with respect $x$ and using the
relation (2.8) we obtain:
\begin{center}
{${\frac {{\partial}}{{\partial}x}}\ {\Bigg {[}}\ {\hbar}\ {\frac
{{\partial}S}{{\partial}t}}\ +\ {\Big {(}}\ {\frac
{1}{2}}\ m\ v_{qu}^{2}\ +\ {\big {[}}\ {\frac {1}{2}}\ m\ {\Omega}^{2}(t)\
x^{2}\ +\ {\lambda}\ x\ X(t)\ {\big {]}}\ +\ V_{qu}\ {\Big
{)}}\ {\Bigg {]}}\ =\ 0\ \ \ {\to}$}
\end{center}
\begin{center}
{${\frac {{\partial}}{{\partial}t}}\ {\big {(}}\ {\frac {{\hbar}}{m}}\
{\frac {{\partial}S}{{\partial}x}}\ {\big {)}}\ +\ {\frac
{{\partial}}{{\partial}x}}\ {\big {(}}\ {\frac {1}{2}}\ v_{qu}^{2}\
{\big {)}}\ +\ {\Omega}^{2}(t)\ x\ +\ {\frac {{\lambda}}{m}}\ X(t)\ =\
-\ {\frac {1}{m}}\ {\frac {{\partial}}{{\partial}x}}\ V_{qu}\ \ \
{\to}$}
\end{center}
\begin{center}
{${\frac {{\partial}v_{qu}}{{\partial}t}}\ +\ v_{qu}\ {\frac
{{\partial}v_{qu}}{{\partial}x}}\ +\ {\Omega}^{2}(t)\ x\ +\ {\frac
{{\lambda}}{m}}\ X(t)\ =\ -\ {\frac {1}{m}}\ {\frac
{{\partial}}{{\partial}x}}\ V_{qu}$\ ,\ \ \ \ \ (2.13)}
\end{center}
which is an equation similar to the Euler's equation which governs the
motion of an ideal fluid.
\par
Taking into account that[6]:
\begin{center}
{$v_{qu}(x,\ t)\ =\ {\frac {dx_{qu}}{dt}}\ =\ {\big {[}}\ {\frac {{\dot
{{\delta}}}(t)}{{\delta}(t)}}\ +\ {\frac {1}{2\ {\tau}}}\ {\big {]}}\
[x_{qu}\ -\ q(t)]\ +\ {\dot {q}}(t)$\ ,\ \ \ \ (2.14a,b)}
\end{center}
the expression (2.13) could be written as:
\begin{center}
{$m\ {\frac {d^{2}x}{dt^{2}}}\ =\ -\ {\frac {{\partial}}{{\partial}x}}\
{\Big {[}}\ {\frac {1}{2}}\ m\ {\Omega}^{2}(t)\ x^{2}\ +\ {\lambda}\ x\
X(t)\ + V_{qu}\ {\Big {]}}\ {\equiv}$}
\end{center}
\begin{center}
{${\equiv}\ F_{c}(x,\ t)\ {\mid}_{x = x(t)}\ +\
F_{qu}(x,\ t)\ {\mid}_{x = x(t)}$\ ,\ \ \ \ \ (2.15)}
\end{center}
where:
\begin{center}
{${\frac {d}{dt}}\ =\ {\frac {{\partial}}{{\partial}t}}\ +\ v_{qu}\
{\frac {{\partial}}{{\partial}x}}$\ ,\ \ \ \ \ (2.16)}
\end{center}
is the "substantive differentiation" (local plus convective) or
"hidrodynamical differention" [5]. We note that the eq. (2.15) has a
form of the second Newton law.
\par
In this way, the expressions (2.10a,b;2.13;2.15) represent the dynamics of
a quantum particle which propagates with the quantum velocity (${\vec
{v}}_{qu}$) in a viscous medium, submitted to a time dependent
classical harmonic potential [${\frac {1}{2}}\ m\ {\Omega}(t)^{2}\
x^{2}$], to an external linear field characterized by the potential
[${\lambda}\ x\ X(t)$] and to the quantum Bohm potential ($V_{qu}$).
\par
In what follows we calculate the wave packet of the Schr\"{o}dinger's
Equation for Continuous Measurements ($SECM$) given by the eq.(2.1)
\vspace{0.2cm}
\par
{\bf 3.\ Quantum Wave Packet}
\vspace{0.2cm}
\par
{\bf 3.1.\ Introduction}
\vspace{0.2cm}
\par
In 1909 [7], Einstein studied the black body radiation in thermodynamical
equilibrium with matter. Starting from Planck's equation, of 1900, of
the radiation density and using the Fourier expansion technique to
calculate its fluctuations, he showed that it exhibits, simultaneously,
fluctuations which are characteristic of waves and particles. In 1916
[8], analyzing again the black body Planckian radiation, Einstein
proposed that an electromagnetic radiation with wavelenght ${\lambda}$
had a linear momentum $p$, given by the relation:
\begin{center}
{p\ =\ ${\frac {h}{{\lambda}}}$\ ,\ \ \ \ \ (3.1.1)}
\end{center}
where {\bf h} is the Planck constant [9].
\par
In works developed between 1923 and 1925 [10] de Broglie formulated his
fundamental idea that the electron with mass $m$, in its atomic orbital
motion with velocity $v$ and linear momentum $p\ =\ m\ v$ is guided by
a "matter wave" (pilot-wave) with wavelenght is given by:
\begin{center}
{${\lambda}\ =\ {\frac {h}{p}}\ .\ \ \ \ \ (3.1.2)$}
\end{center}
\par
In 1926 [11], Schr\"{o}\-dinger proposed that the "pilot-wave de
Brogliean" ought to obey a differential equation, today know as the
famous Schr\"{o}dinger's equation:
\begin{center}
{$i\ {\hbar}\ {\frac {{\partial}}{{\partial}t}}\ {\Psi}({\vec {r}},\ t)
=\ {\hat {H}}\ {\Psi}({\vec {r}},\ t)$\ ,\ \ \ \ \ (3.1.3a)}
\end{center}
where ${\hat {H}}$ is Hamiltonian operator definied by:
\begin{center}
{${\hat {H}}\ =\ {\frac {{\hat {p}}^{2}}{2\ m}}\ +\ V({\vec {r}},\ t)\
,\ \ \ \ \ ({\hat {p}}\ =\ -\ i\ {\hbar}\ {\nabla})$\ ,\ \ \ \ \
(3.1.3b,c)}
\end{center}
where $V$ is the potential energy. In this same year of 1926 [12] Born
interpreted the Schr\"{o}\-dinger wave function ${\Psi}$} as being an
amplitude of probability.
\vspace{0.2cm}
\par
{\bf 3.2.\ Quantum Wave Packet via Schr\"{o}dinger-Feynman Quantum
Mecha\-nics}
\vspace{0.2cm}
\par
As well known [13], when the potential energy $V$ of physical system,
with total energy $E$, depends only on the position [$V({\vec {r}}$)],
the solution of the Schr\"{o}dinger's equation ($SE$) [see relations
(3.1.3a-c)] is given by:
\begin{center}
{${\Psi}({\vec {r}},\ t)\ =\ {\psi}({\vec {r}})\ e^{-\ i\ {\frac
{E}{{\hbar}}}\ t}\ =\ {\psi}({\vec {r}})\ e^{-\ i\ {\omega}\ t}$,\ \ \ \
\ (3.2.1a,b)}
\end{center}
where ${\psi}({\vec {r}}$) satisfies the following equation, known as
time independent $SE$ [14]:
\begin{center}
{${\Delta}\ {\psi}({\vec {r}})\ +\ {\frac {2\ m}{{\hbar}^{2}}}\ (E\ -\
V)\ {\psi}({\vec {r}})\ =\ 0\ \ \ {\Leftrightarrow}\ \ \ {\hat {H}}\
{\psi}({\vec {r}})\ =\ E\ {\psi}({\vec {r}})$,\ \ \ \ \ (3.2.2a,b)}
\end{center}
where ${\hat {H}}$ is given by the expressions (3.1.3b,c). In addition,
${\psi}({\vec {r}}$) and its differentiation must be continuous ${\frac
{{\partial}{\psi}({\vec {r}})}{{\partial}{\vec {r}}}}$.
\par
It is important to note that the relation (3.2.1b) was obtained
considering the Planckian energy:
\begin{center}
{$E\ =\ h\ {\nu}\ =\ {\hbar}\ {\omega}\ ,\ \ \ {\hbar}\ =\ {\frac
{h}{2\ {\pi}}},\ \ \ {\omega}\ =\ 2\ {\pi}\ {\nu}$\ .\ \ \ \ \
(3.2.3a-d)}
\end{center}
\par
As the expression (3.2.2b) is an eigenvalue equation its solution is
given by a discrete set of eigenfunctions ("Schr\"{o}dingerian waves")
of the operator ${\hat {H}}$ [15]. On the other hand, the expression
(3.1.2) suggests that it would be possible to use a handful
concentration of "de Brogliean waves", with wavelenght ${\lambda}$, to
describe the particles localized in the space. In this description it
is necessary to use a mechanism that takes into account these "waves"
with many wavelenghts. This mechanism is the Fourier Analysis [15]. So,
according to this technique (to one dimentional case) we can considerer
${\psi}$(x) like a superposition of plane monochromatic harmonic waves,
that is:
\begin{center}
{${\psi}(x)\ =\ {\frac {1}{{\sqrt {2\ {\pi}}}}}\ {\int}_{-\
{\infty}}^{+\ {\infty}}\ {\phi}(k)\ e^{i\ k\ x}\ dk$\ ,\ \ \ \ \
(3.2.4a)}
\end{center}
beeing:
\begin{center}
{${\phi}(k)\ =\ {\frac {1}{{\sqrt {2\ {\pi}}}}}\ {\int}_{-\
{\infty}}^{+\ {\infty}}\ {\psi}(x')\ e^{-\ i\ k\ x'}\ dx'$\ ,\ \ \ \ \
(3.2.4b)}
\end{center}
where:
\begin{center}
{$k\ =\ {\frac {2\ {\pi}}{{\lambda}}}$\ ,\ \ \ \ \ (3.2.4c)}
\end{center}
is the wavenumber and represents the transition from the discrete to
the continuous description. Note that, using the expressions (3.1.2),
(3.2.3c), the relation (3.2.4c) will be written as:
\begin{center}
{$k\ =\ 2\ {\pi}\ {\frac {p}{h}}\ \ \ {\to}\ \ \ p\ =\ {\hbar}\ k$\ .\
\ \ \ \ (3.2.4d)}
\end{center}
\par
Inserting the expression (3.2.4a) in the same relation (3.2.4b)
results:
\begin{center}
{${\psi}(x)\ =\ {\frac {1}{2\ {\pi}}}\ {\int}_{-\
{\infty}}^{+\ {\infty}}\ {\int}_{-\ {\infty}}^{+\ {\infty}}\
{\psi}(x')\ e^{i\ k\ (x -\ x')}\ dx'\ dk$\ .\ \ \ \ \ (3.2.5)}
\end{center}
\par
Considering that [15]:
\begin{center}
{${\delta}(z'\ -\ z)\ {\equiv}\ {\delta}(z\ -\ z')\ =\ {\frac {1}{2\
{\pi}}}\ {\int}_{-\ {\infty}}^{+\ {\infty}}\ e^{i\ k\ (z -\ z')}\ dk$\
,\ \ \ \ \ (3.2.6a,b)}
\end{center}
\begin{center}
{$f(z)\ =\ {\int}_{-\ {\infty}}^{+\ {\infty}}\ f(z')\ {\delta}(z'\ -\
z)\ dz'$\ ,\ \ \ \ \ (3.2.6c)}
\end{center}
we verify that (since  $z,\ z'\ {\equiv}\ x,\ x'$), the consistency of
the relation (3.2.5), characterized by the famous completeness
relation.
\par
Taking into account the relation (3.2.4a) in the unidimensional
representation of the equation (3.2.1b) the following relation will be
obtained:
\begin{center}
{${\Psi}(x,\ t)\ =\ {\frac {1}{{\sqrt {2\ {\pi}}}}}\ {\int}_{-\
{\infty}}^{+\ {\infty}}\ {\phi}(k)\ e^{i\ [k\ x\ -\ {\omega}(k)\ t]}\
dk$\ ,\ \ \ \ \ (3.2.7)}
\end{center}
which represents the wave packet of amplitude ${\phi}(k)$.
\par
Note that the dependence ${\omega}$ in terms of $k$, indicated in the
above relation, is due to the fact that the energy $E$ of a physical
system depends of $p$. So, considering that this fact and also the
relations (3.2.3b;3.2.4d), this dependence is verified immediately as will
be shown in what follows.
\par
Now, let us write the equation (3.2.7) in terms of the Feynman
propagator. Putting $t\ =\ 0$ in this expression and, in analogy with
the relation (3.2.4b), we get:
\begin{center}
{${\Psi}(x,\ 0)\ =\ {\frac {1}{{\sqrt {2\ {\pi}}}}}\ {\int}_{-\
{\infty}}^{+\ {\infty}}\ {\phi}(k)\ e^{i\ k\ x}\ dk\ \ \ {\to}$}
\end{center}
\begin{center}
{${\phi}(k)\ =\ {\frac {1}{{\sqrt {2\ {\pi}}}}}\ {\int}_{-\ {\infty}}^{+\
{\infty}}\ {\Psi}(x',\ 0)\ e^{-\ i\ k\ x'}\ dx'$\ .\ \ \ \ \ (3.2.8)}
\end{center}
\par
Inserting this expression in the equation (3.2.7) results:
\begin{center}
{${\Psi}(x,\ t)\ =\ {\frac {1}{{\sqrt {2\ {\pi}}}}}\ {\int}_{-\
{\infty}}^{+\ {\infty}}\ {\Big {(}}\ {\frac {1}{{\sqrt {2\ {\pi}}}}}\
{\int}_{-\ {\infty}}^{+\ {\infty}}\ {\Psi}(x',\ 0)\ e^{-\ i\ k\ x'}\
dx'\ {\Big {)}}\ {\times}\ e^{i\ [k\ x\ -\ {\omega}(k)\ t]}\
dk\ \ \ {\to}$}
\end{center}
\begin{center}
{${\Psi}(x,\ t)\ =\ {\int}_{-\ {\infty}}^{+\ {\infty}}\ {\Big {(}}\
{\frac {1}{2\ {\pi}}}\ {\int}_{-\ {\infty}}^{+\ {\infty}}\ e^{i\ k\
[(x\ -\ x')\ -\ {\frac {{\omega}(k)}{k}}\ t]}\ dk\ {\Big {)}}\
{\times}\ {\Psi}(x',\ 0)\ dx'$\ .\ \ \ \ \ (3.2.9)}
\end{center}
\par
At this point it is important to say that, in the formalism of Feynman
Quantum Mechanics [16], the term inside the brackets in the equation
(3.2.9) represents the Feynman propagator $K(x,\ x';\ t)$. So, this
expression can be written as:
\begin{center}
{${\Psi}(x,\ t)\ =\ {\int}_{-\ {\infty}}^{+\ {\infty}}\ K(x,\ x';\
t)\ {\Psi}(x',\ 0)\ dx'$\ ,\ \ \ \ \ (3.2.10a)}
\end{center}
where:
\begin{center}
{$K(x,\ x';\ t)\ =\ {\frac {1}{2\ {\pi}}}\ {\int}_{-\ {\infty}}^{+\
{\infty}}\ e^{i\ k\ [(x\ -\ x')\ -\ {\frac {{\omega}(k)}{k}}\ t]}\ dk$\
.\ \ \ \ \ (3.2.10b)}
\end{center}
\par
The equation (3.2.10a) represents the wavefunction ${\Psi}$ for any
time $t$ in terms of this function in the time $t\ =\ 0$. So, if
${\omega}$(k) is a known function of $k$, so ${\Psi}(x,\ t)$ can be
explicitely obtained from ${\Psi}(x,\ 0)$.
\par
In the sequence we determine the form of the wave packet given in the
equations (3.2.10a,b) according to the $SECM$, defined by eq.(2.1) [17].
\vspace{0.2cm}
\par
{\bf 3.3.\ The Quantum Wave Packet of the Schr\"{o}dinger for
Continuous Measurements}
\vspace{0.2cm}
\par
Initially, let us calculate the quantum trajectory ($x_{qu}$) of the
physical system represented by the eq.(2.1). To do this, let us
integrate the relations (2.14b) [remembering that ${\int}\ {\frac
{dz}{z}}\ =\ {\ell}n\ z,\ {\ell}n\ ({\frac {x}{y}})\ =\ {\ell}n\ x\ -\
{\ell}n\ y$\ ,e\ ${\ell}n\ x\ y\ =\ {\ell}n\ x\ +\ {\ell}n\ y$]:
\begin{center}
{$v_{qu}(x,\ t)\ =\ {\frac {dx_{qu}}{dt}}\ =\ {\big {[}}\ {\frac {{\dot
{{\delta}}}(t)}{{\delta}(t)}}\ +\ {\frac {1}{2\ {\tau}}}\ {\big {]}}\
[x_{qu}\ -\ q(t)]\ +\ {\dot {q}}(t)\ \ \ {\to}$}
\end{center}
\begin{center}
{${\frac {dx_{qu}}{dt}}\ -\ {\frac {dq}{dt}}\ =\ {\big {[}}\ {\frac {{\dot
{{\delta}}}(t)}{{\delta}(t)}}\ +\ {\frac {1}{2\ {\tau}}}\ {\big {]}}\
[x_{qu}\ -\ q(t)]\ \ \ {\to}\ \ \ {\frac {d[x_{qu}(t)\ -\
q(t)]}{[x_{qu}(t)\ -\ q(t)]}}\ =\ {\big {[}}\ {\frac {{\dot
{{\delta}}}(t)}{{\delta}(t)}}\ dt\ +\ {\frac {dt}{2\ {\tau}}}\ {\big
{]}}\ \ \ {\to}$}
\end{center}
\begin{center}
{${\int}_{o}^{t}\ {\frac {d[x_{qu}(t')\ -\ q(t')]}{[x_{qu}(t')\ -\
q(t')]}}\ =\ {\int}_{o}^{t}\ {\frac {d{\delta}(t')}{{\delta}(t')}}\ +\
{\int}_{o}^{t}\ {\frac {dt}{2\ {\tau}}}\ \ \ {\to}$}
\end{center}
\begin{center}
{${\ell}n\ {\Big {(}}\ {\frac {[x_{qu}(t)\ -\ q(t)]}{[x_{qu}(0)\ -\
q(0)]}}\ {\Big {)}}\ =\ {\ell}n\ {\Big {[}}\ {\frac
{{\delta}(t)}{{\delta}(0)}}\ {\Big {]}}\ +\ {\frac {t}{2\ {\tau}}}\ =\
{\ell}n\ {\Big {[}}\ {\frac {{\delta}(t)}{{\delta}(0)}}\ {\Big {]}}\ +\
{\ell}n\ {\Big {[}}\ exp\ {\big {(}}\ {\frac {t}{2\ {\tau}}}\ {\big
{)}}\ {\Big {]}}\ =$}
\end{center}
\begin{center}
{$\ =\ {\ell}n\ {\Big {(}}\ {\frac {{\delta}(t)}{{\delta}(0)}}\ .\ exp\
{\Big {[}}\ {\frac {t}{2\ {\tau}}}\ {\Big {]}}\ {\Big {)}}\
\ \ {\to}\ \ \ x_{qu}(t)\ =\ q(t)\ +\ e^{t/2\ {\tau}}\ {\frac
{{\delta}(t)}{{\delta}(0)}}\ [x_{qu}(0)\ -\ q(0)]$\ ,\ \ \ \ \ (3.3.1)}
\end{center}
that represent the looked for quantum trajectory.
\par
To obtain the Schr\"{o}\-dinger-de Broglie-Bohm wave packet for
Continuous Measurements given by the eq.(2.2), let us expand the
functions $S(x,\ t)$, $V(x,\ t)$ and $V_{qu}(x,\ t)$ around of $q(t)$
up to second Taylor order [2.5]. In this way we have:
\begin{center}
{$S(x,\ t)\ =\ S[q(t),\ t]\ +\ S'[q(t),\ t]\ [x\ -\ q(t)]\ +\ {\frac
{S''[q(t),\ t]}{2}}\ [x\ -\ q(t)]^{2}$\ ,\ \ \ \ \ (3.3.2)}
\end{center}
\begin{center}
{$V(x,\ t)\ =\ V[q(t),\ t]\ +\ V'[q(t),\ t]\ [x\ -\ q(t)]\ +\ {\frac
{V''[q(t),\ t]}{2}}\ [x\ -\ q(t)]^{2}$\ ,\ \ \ \ \ (3.3.3)}
\end{center}
\begin{center}
{$V_{qu}(x,\ t)\ =\ V_{qu}[q(t),\ t]\ +\ V_{qu}'[q(t),\ t]\ [x\ -\
q(t)]\ +\ {\frac {V_{qu}''[q(t),\ t]}{2}}\ [x\ -\ q(t)]^{2}$\ .\ \ \ \
\ (3.3.4)}
\end{center}
\par
Differentiating the expression (3.3.2) in the variable $x$, multiplying
the result by ${\frac {{\hbar}}{m}}$, using the relations (2.8) and
(2.14b), and taking into account the polynomial identity property, we
obtain:
\begin{center}
{${\frac {{\hbar}}{m}}\ {\frac {{\partial}S(x,\ t)}{{\partial}x}}\ =\
{\frac {{\hbar}}{m}}\ {\Big {(}}\ S'[q(t),\ t]\ +\ S''[q(t),\ t]\ [x\
-\ q(t)]\ {\Big {)}}\ =$}
\end{center}
\begin{center}
{$=\ v_{qu}(x,\ t)\ =\ {\big {[}}\ {\frac {{\dot
{{\delta}}}(t)}{{\delta}(t)}}\ +\ {\frac {1}{2\ {\tau}}}\ {\big {]}}\
[x_{qu}\ -\ q(t)]\ +\ {\dot {q}}(t)\ =\ \ \ {\to}$}
\end{center}
\begin{center}
{$S'[q(t),\ t]\ =\ {\frac {m\ {\dot {q}}(t)}{{\hbar}}}\ ,\ \ \
S''[q(t),\ t]\ =\ {\frac {m} {{\hbar}}}\ {\big {[}}\ {\frac {{\dot
{{\delta}}}(t)}{{\delta}(t)}}\ +\ {\frac {1}{2\ {\tau}}}\ {\big {]}}$\
.\ \ \ \ \ (3.3.5a,b)}
\end{center}
\par
Substituting the expressions (3.3.5a,b) in the equation (3.3.2),
results:
\begin{center}
{$S(x,\ t)\ =\ S_{o}(t)\ +\ {\frac {m\ {\dot {q}}(t)}{{\hbar}}}\ [x\ -\
q(t)]\ +\ {\frac {m}{2\ {\hbar}}}\ {\Big {[}}\ {\frac {{\dot
{{\delta}}}(t)}{{\delta}(t)}}\ +\ {\frac {1}{2\ {\tau}}}\ {\Big {]}}\
[x\ -\ q(t)]^{2}$\ ,\ \ \ \ \ (3.3.6)}
\end{center}
where:
\begin{center}
{$S_{o}(t)\ {\equiv}\ S[q(t),\ t]$\ ,\ \ \ \ \ (3.3.7)}
\end{center}
is the quantum action.
\par
Differentiating the eq.(3.3.6) in relation to the time $t$, we obtain
(remembering that ${\frac {{\partial}x}{{\partial}t}}$\ =\ 0):
\begin{center}
{${\frac {{\partial}S}{{\partial}t}}\ =\ {\dot {S}}_{o}(t)\ +\ {\frac
{{\partial}}{{\partial}t}}\ {\Big {(}}\ {\frac {m\ {\dot
{q}}(t)}{{\hbar}}}\ [x\ -\ q(t)]\ {\Big {)}}\ +\ {\frac
{{\partial}}{{\partial}t}}\ {\Bigg {(}}\ {\frac {m}{2\ {\hbar}}}\ {\Big
{[}}\ {\frac {{\dot {{\delta}}}(t)}{{\delta}(t)}}\ +\ {\frac {1}{2\
{\tau}}}\ {\Big {]}}\ [x\ -\ q(t)]^{2}\ {\Bigg {)}}\ \ \ {\to}$}
\end{center}
\begin{center}
{${\frac {{\partial}S}{{\partial}t}}\ =\ {\dot {S}}_{o}(t)\ +\ {\frac
{m\ {\ddot {q}}(t)}{{\hbar}}}\ [x\ -\ q(t)]\ -\ {\frac {m\ {\dot
{q}}(t)^{2}}{{\hbar}}}\ +$}
\end{center}
\begin{center}
{+\ ${\frac {m}{2\ {\hbar}}}\ [{\frac {{\ddot
{{\delta}}}(t)}{{\delta}(t)}}\ -\ {\frac {{\dot
{{\delta}}}^{2}(t)}{{\delta}^{2}(t)}}]\ [x\ -\ q(t)]^{2}\ -\ {\frac {m\
{\dot {q}}(t)}{{\hbar}}}\ {\Big {(}}\ {\frac {{\dot
{{\delta}}}(t)}{{\delta}(t)}}\ +\ {\frac {1}{2\ {\tau}}}\ {\Big {)}}\
[x\ -\ q(t)]$\ .\ \ \ \ \ (3.3.8)}
\end{center}
\par
Considering that [6]:
\begin{center}
{${\rho}(x,\ t)\ =\ [2\ {\pi}\ {\delta}^{2}(t)]^{-\ 1/2}\ e^{-\ {\frac {[x\ -\
{\bar {x}}(t)]^{2}}{2\ {\delta}^{2}(t)}}}$\ ,\ \ \ \ \ \ (3.3.9)}
\end{center}
let us write $V_{qu}$ in terms of $[x\ -\ q(t)]$. Initially using
the eqs.(2.7b) and (3.3.9), we calculate the following
differentiations:
\begin{center}
{${\frac {{\partial}{\phi}}{{\partial}x}}\ =\ {\frac
{{\partial}}{{\partial}x}}\ {\Big {(}}\ [2\ {\pi}\ {\delta}^{2}(t)]^{-\
1/4}\ e^{-\ {\frac {[x\ -\ q(t)]^{2}}{4\ {\delta}^{2}(t)}}}\ {\Big
{)}}\ =\ [2\ {\pi}\ {\delta}^{2}(t)]^{-\ 1/4}\ e^{-\ {\frac {[x\ -\
q(t)]^{2}}{4\ {\delta}^{2}(t)}}} {\frac {{\partial}}{{\partial}x}}\
{\Big {(}}\ -\ {\frac {[x\ -\ q(t)]^{2}}{4\ {\delta}^{2}(t)}}\ {\Big
{)}}\ \ \ {\to}$}
\end{center}
\begin{center}
{${\frac {{\partial}{\phi}}{{\partial}x}}\ =\ -\ [2\ {\pi}\
{\delta}^{2}(t)]^{-\ 1/4}\ e^{-\ {\frac {[x\ -\ q(t)]^{2}}{4\
{\delta}^{2}(t)}}}\ {\frac {[x\ -\ q(t)]}{2\ {\delta}^{2}(t)}}$\ ,}
\end{center}
\begin{center}
{${\frac {{\partial}^{2}{\phi}}{{\partial}x^{2}}}\ =\ {\frac
{{\partial}}{{\partial}x}}\ {\Big {(}}\ -\ [2\ {\pi}\
{\delta}^{2}(t)]^{-\ 1/4}\ e^{-\ {\frac {[x\ -\ q(t)]^{2}}{4\
{\delta}^{2}(t)}}}\ {\frac {[x\ -\ q(t)]}{2\ {\delta}^{2}(t)}}\ {\Big
{)}}$\ =}
\end{center}
\begin{center}
{$\ =\ -\ [2\ {\pi}\ {\delta}^{2}(t)]^{-\ 1/4}\ e^{-\ {\frac {[x\ -\
q(t)]^{2}}{4\ {\delta}^{2}(t)}}}\ {\frac {{\partial}}{{\partial}x}}\
{\Big {(}}\ {\frac {[x\ -\ q(t)]}{2\ {\delta}^{2}(t)}}\ {\Big {)}}\ -$}
\end{center}
\begin{center}
{$-\ [2\ {\pi}\ {\delta}^{2}(t)]^{-\ 1/4}\ e^{-\ {\frac {[x\ -\
q(t)]^{2}}{4\ {\delta}^{2}(t)}}}\ {\frac {{\partial}}{{\partial}x}}\
{\Big {(}}\ -\ {\frac {[x\ -\ q(t)]^{2}}{4\ {\delta}^{2}(t)}}\ {\Big
{)}}\ {\big {(}}\ {\frac {[x\ -\ q(t)]}{2\ {\delta}^{2}(t)}}\ {\big
{)}}\ \ \ {\to}$}
\end{center}
\begin{center}
{${\frac {{\partial}^{2}{\phi}}{{\partial}x^{2}}}\ =\ -\ [2\ {\pi}\
{\delta}^{2}(t)]^{-\ 1/4}\ e^{-\ {\frac {[x\ -\ q(t)]^{2}}{4\
{\delta}^{2}(t)}}}\ {\frac {1}{2\ {\delta}^{2}(t)}}\ +\ [2\ {\pi}\
{\delta}^{2}(t)]^{-\ 1/4}\ e^{-\ {\frac {[x\ -\ q(t)]^{2}}{4\
{\delta}^{2}(t)}}}\ {\frac {[x\ -\ q(t)]^{2}}{4\ {\delta}^{4}(t)}}$\ =}
\end{center}
\begin{center}
{$=\ -\ {\phi}\ {\frac {1}{2\ {\delta}^{2}(t)}}\ +\ {\phi}\ {\frac {[x\
-\ q(t)]^{2}}{4\ {\delta}^{4}(t)}}\ \ \ {\to}\ \ \ {\frac {1}{{\phi}}}\
{\frac {{\partial}^{2}{\phi}}{{\partial}x^{2}}}\ =\ {\frac {[x\ -\
q(t)]^{2}}{4\ {\delta}^{4}(t)}}\ -\ {\frac {1}{2\ {\delta}^{2}(t)}}$\
.\ \ \ \ \ (3.3.10)}
\end{center}
\par
Substituting the relation (3.3.10) in the equation (2.10a), taking into
account the expression (3.3.4), results:
\begin{center}
{$V_{qu}(x,\ t)\ =\ V_{qu}[q(t),\ t]\ +\ V_{qu}'[q(t),\ t]\ [x\ -\
q(t)]\ +\ {\frac {V_{qu}''[q(t),\ t]}{2}}\ [x\ -\ q(t)]^{2}\ \ \ {\to}$}
\end{center}
\begin{center}
{$V_{qu}(x,\ t)\ =\ {\frac {{\hbar}^{2}}{4\ m\ {\delta}^{2}(t)}}\ [x\
-\ q(t)]^{o}\ -\ {\frac {{\hbar}^{2}}{8\ m\ {\delta}^{4}(t)}}\ [x\ -\
q(t)]^{2}$\ .\ \ \ \ \ (3.3.11)}
\end{center}
\par
Besides this the eq.(3.3.3) will be written, using the eq.(2.1) in the
form:
\begin{center}
{$V(x,\ t)\ =\ V[q(t),\ t]\ +\ V'[q(t),\ t]\ [x\ -\ q(t)]\ +\ {\frac
{V''[q(t),\ t]}{2}}\ [x\ -\ q(t)]^{2}\ \ \ {\to}$}
\end{center}
\begin{center}
{$V(x,\ t)\ =\ {\frac {1}{2}}\ m\ {\Omega}^{2}(t)\ q^{2}(t)\ +\
{\lambda}\ q(t)\ X(t)\ +$}
\end{center}
\begin{center}
{$+\ {\Big {(}}\ m\ {\Omega}^{2}(t)\ q(t)\ +\ {\lambda}\ X(t)\ {\Big
{)}}\ [x\ -\ q(t)]\ +\ {\frac {m}{2}}\ {\Omega}^{2}(t)\ [x\ -\
q(t)]^{2}$\ .\ \ \ \ \ (3.3.12)}
\end{center}
\par
Inserting the relations (2.8), (2.14b) and (3.3.2-4;3.3.8,10,11), into the
eq.(2.12b), we obtain, remembering that $S_{o}(t)$, ${\delta}(t)$ and
$q(t)$:
\begin{center}
{${\hbar}\ {\frac {{\partial}S}{{\partial}t}}\ +\ {\frac {1}{2}}\ m\
v_{qu}^{2}\ +\ V\ +\ V_{qu}\ =$}
\end{center}
\begin{center}
{$=\ {\hbar}\ {\Big {[}}\ {\dot {S}}_{o}\ +\ {\frac {m\ {\ddot
{q}}}{{\hbar}}}\ (x\ -\ q)\ -\ {\frac {m\ {\dot
{q}}^{2}}{{\hbar}}}\ +\ {\frac {m}{2\ {\hbar}}}\ {\big {(}}\ {\frac
{{\ddot {{\delta}}}}{{\delta}}}\ -\ {\frac {{\dot
{{\delta}}}^{2}}{{\delta}^{2}}}\ {\big {)}}\ (x\ -\ q)^{2}\ -$}
\end{center}
\begin{center}
{$-\ {\frac {m\ {\dot {q}}}{{\hbar}}}\ {\big {(}}\ {\frac {{\dot
{{\delta}}}}{{\delta}}}\ +\ {\frac {1}{2\ {\tau}}}\ {\big {)}}\ (x\ -\
q) {\Big {]}}\ +\ {\frac {1}{2}}\ m\ {\Big {[}}\ {\big {(}}\ {\frac
{{\dot {{\delta}}}}{{\delta}}}\ +\ {\frac {1}{2\ {\tau}}}\ {\big {)}}\
(x\ -\ q)\ +\ {\dot {q}}\ {\Big {]}}^{2}\ +$}
\end{center}
\begin{center}
{$+\ {\frac {1}{2}}\ m\ {\Omega}^{2}(t)\ q^{2} +\
{\lambda}\ q\ X(t)\ +\ {\big {[}}\ m\ {\Omega}^{2}(t)\ q\ +\
{\lambda}\ X(t)\ {\big {]}}\ (x\ -\ q)\ +\ {\frac {m}{2}}\
{\Omega}^{2}(t)\ (x\ -\ q)^{2}\ +$}
\end{center}
\begin{center}
{$+\ {\frac {{\hbar}^{2}}{4\ m\ {\delta}^{2}}}\ -\ {\frac
{{\hbar}^{2}}{8\ m\ {\delta}^{4}}}\ (x\ -\ q)^{2}\ =\ 0$\ .\ \ \ \ \
(3.3.13)}
\end{center}
\par
Since $(x\ -\ q)^{o}\ =\ 1$, we can gather together the above
expression in potencies of $(x\ -\ q)$, obtaining:
\begin{center}
{${\Big {[}}\ {\hbar}\ {\dot {S}}_{o}\ -\ m\ {\dot {q}}^{2}\ +\
{\frac {1}{2}}\ m\ {\dot {q}}^{2}\ +\ {\frac {1}{2}}\ m\
{\Omega}^{2}(t)\ q^{2} +\ {\lambda}\ q\ X(t)\ +\ {\frac
{{\hbar}^{2}}{4\ m\ {\delta}^{2}}}\ {\Big {]}}\ (x\ -\ q)^{o}\ +$}
\end{center}
\begin{center}
{$+\ {\Big {[}}\ m\ {\ddot {q}}\ -\ m\ {\dot {q}}\ {\big {(}}\ {\frac
{{\dot {{\delta}}}}{{\delta}}}\ +\ {\frac {1}{2\ {\tau}}}\ {\big {)}}\
+\ m\ {\dot {q}}\ {\big {(}}\ {\frac {{\dot {{\delta}}}}{{\delta}}}\ +\
{\frac {1}{2\ {\tau}}}\ {\big {)}} +\ m\ {\Omega}^{2}(t)\ q\ +\
{\lambda}\ X(t)\ {\Big {]}}\ (x\ -\ q)\ +\ {\Big {[}}\ {\frac {m}{2}}\
{\big {(}}\ {\frac {{\ddot {{\delta}}}}{{\delta}}}\ -\ {\frac {{\dot
{{\delta}}}^{2}}{{\delta}^{2}}}\ {\big {)}}\ +$}
\end{center}
\begin{center}
{$+\ {\frac {m}{2}}\ {\big {(}}\ {\frac {{\dot
{{\delta}}}^{2}}{{\delta}^{2}}}\ +\ {\frac {{\dot {{\delta}}}}{{\tau}\
{\delta}}}\ +\ {\frac {1}{4\ {\tau}^{2}}}\ {\big {)}}\ +\ {\frac {m}{2}}\
{\Omega}^{2}(t)\ -\ {\frac {{\hbar}^{2}}{8\ m\ {\delta}^{4}}}\ {\Big
{]}}\ (x\ -\ q)^{2}\ =\ 0$\ .\ \ \ \ \ (3.3.14)}
\end{center}
\par
As the above relation is an identically null polynomium, the
coefficients of the potencies must be all equal to zero, that is:
\begin{center}
{${\dot {S}}_{o}(t)\ =\ {\frac {1}{{\hbar}}}\ {\Big {[}}\ {\frac
{1}{2}}\ m\ {\dot {q}}^{2}\ -\ {\frac {1}{2}}\ m\ {\Omega}^{2}(t)\
q^{2} -\ {\lambda}\ q\ X(t)\ -\ {\frac {{\hbar}^{2}}{4\ m\
{\delta}^{2}}}\ {\Big {]}}$\ ,\ \ \ \ \ (3.3.15)}
\end{center}
\begin{center}
{${\ddot {q}}\ +\ {\Omega}^{2}(t)\ q\ +\ {\frac {{\lambda}}{m}}\
X(t)\ =\ 0$\ ,\ \ \ \ \ (3.3.16)}
\end{center}
\begin{center}
{${\ddot {{\delta}}}\ +\ {\frac {{\dot {{\delta}}}}{{\tau}}}\ +\ {\Big
{[}}\ {\Omega}^{2}(t)\ +\ {\frac {1}{4\ {\tau}^{2}}}\ {\Big {]}}\
{\delta}\ =\ {\frac {{\hbar}^{2}}{4\ m^{2}\ {\delta}^{3}(t)}}$\ .\ \ \ \ \
(3.3.17)}
\end{center}
\par
Assuming that the following initial conditions are obeyed:
\begin{center}
{$q(0)\ =\ x_{o}\ ,\ \ \ {\dot {q}}(0)\ =\ v_{o}\ ,\ \ \ {\delta}(0)\
=\ a_{o}\ ,\ \ \ {\dot {{\delta}}}(0)\ =\ b_{o}$\ ,\ \ \ \ \ \ (3.3.18a-d)}
\end{center}
and that [see eq.(3.3.7)]:
\begin{center}
{$S_{o}(0)\ =\ {\frac {m\ v_{o}\ x_{o}}{{\hbar}}}$\ ,\ \ \ \ \ (3.3.19)}
\end{center}
the integration of the expression (3.3.15) will be given by:
\begin{center}
{$S_{o}(t)\ =\ {\frac {1}{{\hbar}}}\ {\int}_{o}^{t}\ dt'\ {\Big {[}}\
{\frac {1}{2}}\ m\ {\dot {q}}^{2}(t')\ -\ {\frac {1}{2}}\ m\
{\Omega}^{2}(t')\ q^{2}(t') -$}
\end{center}
\begin{center}
{$-\ {\lambda}\ q(t')\ X(t')\ -\ {\frac {{\hbar}^{2}}{4\ m\
{\delta}^{2}(t')}}\ {\Big {]}}\ +\ {\frac {m\ v_{o}\ x_{o}}{{\hbar}}}$\
.\ \ \ \ \ (3.3.20)}
\end{center}
\par
Taking into account the expressions (3.3.5a,b) and (3.3.20) in the
equation (3.3.6) results:
\begin{center}
{$S(x,\ t)\ =\ {\frac {1}{{\hbar}}}\ {\int}_{o}^{t}\ dt'\ {\Big {[}}\
{\frac {1}{2}}\ m\ {\dot {q}}^{2}(t')\ -\ {\frac {1}{2}}\ m\
{\Omega}^{2}(t')\ q^{2}(t')\ -\ {\lambda}\ q(t')\ X(t')\ -\ {\frac
{{\hbar}^{2}}{4\ m\ {\delta}^{2}(t')}}\ {\Big {]}}\ +$}
\end{center}
\begin{center}
{$+\ {\frac {m\ v_{o}\ x_{o}}{{\hbar}}}\ +\ {\frac {m\ {\dot
{q}}(t)}{{\hbar}}}\ [x\ -\ q(t)]\ +\ {\frac {m}{2\ {\hbar}}}\ {\Big
{[}}\ {\frac {{\dot {{\delta}}}(t)}{{\delta(t)}}}\ +\ {\frac {1}{2\
{\tau}}}\ {\Big {]}}\ [x\ -\ q(t)]^{2}$\ .\ \ \ \ \ (3.3.21)}
\end{center}
\par
This result obtained above permit us, finally, to obtain the wave
packet for the $SBBMC$ equation. Indeed, considering the relations
(2.2;2.7b), (3.3.9) and (3.3.21), we get [18]:
\begin{center}
{${\Psi}(x,\ t)\ =\ [2\ {\pi}\ {\delta}^{2}(t)]^{-\ 1/4}\ exp\ {\Bigg
{[}}\ {\Big {(}}\ {\frac {i\ m}{2\ {\hbar}}}\ {\Big {[}}\ {\frac {{\dot
{{\delta}}}(t)}{{\delta}(t)}}\ +\ {\frac {1}{2\ {\tau}}}\ {\Big {]}}\
-\ {\frac {1}{4\ {\delta}^{2}(t)}}\ {\Big {)}}\ [x\ -\ q(t)]^{2}\ {\Bigg
{]}}\ {\times}$}
\end{center}
\begin{center}
{${\times}\ exp\ {\Big {[}}\ {\frac {i\ m\ {\dot {q}}(t)}{{\hbar}}}\ [x\
-\ q(t)]\ +\ {\frac {i\ m\ v_{o}\ x_{o}}{{\hbar}}}\ {\Big {]}}\ {\times}$}
\end{center}
\begin{center}
{${\times}\ exp\ {\Big {[}}\ {\frac {i}{{\hbar}}}\ {\int}_{o}^{t}\ dt'\
{\Big {[}}\ {\frac {1}{2}}\ m\ {\dot {q}}^{2}(t')\ -\ {\frac {1}{2}}\
m\ {\Omega}^{2}(t')\ q^{2}(t') -\ {\lambda}\ q(t')\ X(t')\ -\ {\frac
{{\hbar}^{2}}{4\ m\ {\delta}^{2}(t')}}\ {\Big {]}}$\ . \ \ \ (3.3.22)}
\end{center}
\vspace{0.2cm}
\begin{center}
{{\bf {\underline {NOTES AND REFERENCES}}}}
\end{center}
\vspace{0.2cm}
\par
1.\ NASSAR, A. B. 2004. {\bf Chaotic Behavior of a Wave Packet under
Continuos Quantum Mechanics} (mimeo).
\par
2.\ To a formal and philosophical study of the $BBQM$ see, for
instance:
\par
2.1. HOLLAND, P. R. 1993. {\bf The Quantum Theory of Motion: An
Account of the de Broglie-Bohm Causal Interpretation of Quantum
Mechanics}, Cambridge University Press.
\par
2.2.\ JAMMER, M. 1974. {\bf The Philosophy of Quantum Mecha\-nics},
John Willey.
\par
2.3.\ FREIRE JUNIOR, O. 1999. {\bf David Bohm e a Controv\'ersia
dos Quanta}, {\it Cole\c{c}\~ao CLE}, Volume 27, Centro de L\'ogica,
Epistemologia e Hist\'oria da Ci\^encia, UNICAMP.
\par
2.4.\ AULETTA, G. 2001. {\bf Foundations and Interpretation of Quantum
Mechanics}, World Scientific.
\par
2.5.\ BASSALO, J. M. F., ALENCAR, P. T. S., CATTANI, M. S. D. e NASSAR,
A. B. 2003. {\bf T\'opicos da Mec\^anica Qu\^antica de de
Broglie-Bohm}, EDUFPA.
\par
3.\ MADELUNG, E. 1926. {\it Zeitschrift f\"{u}r Physik} {\bf 40},
p. 322.
\par
4.\ BOHM, D. 1952. {\it Physical Review} {\bf 85}, p. 166.
\par
5.\ See books on the Fluid Mechanics, for instance:
\par
5.1.\ STREETER, V. L. and DEBLER, W. R. 1966. {\bf Fluid Mechanics},
McGraw-Hill Book Company, Incorporation.
\par
5.2. \ COIMBRA, A. L. 1967. {\bf Mec\^anica dos Meios Cont\'{\i}\-nuos},
Ao Livro T\'ecnico S. A.
\par
5.3.\ LANDAU, L. et LIFSHITZ, E. 1969. {\bf M\'ecanique des Fluides}.
\'Editions Mir.
\par
5.4.\ BASSALO, J. M. F. 1973. {\bf Introdu\c{c}\~ao \`a Mec\^anica
dos Meios Cont\'{\i}nuos}, EDUFPA.
\par
5.5.\ CATTANI, M. S. D. 1990/2005. {\bf Elementos de Mec\^anica dos
Fluidos}, Edgard Bl\"{u}cher.
\par
6.\ BASSALO, J. M. F., ALENCAR, P. T. S., SILVA, D. G. da, NASSAR, A.
B. and CATTANI, M. 2009. {\it arXiv:0902.2988v1}\ {\bf [math-ph]}, 17
February.
\par
7.\ EINSTEIN, A. 1909. {\it Physikalische Zeitschrift} {\bf 10}, p. 185.
\par
8.\ EINSTEIN, A. 1916. {\it Verhandlungen der Deutschen Physikalischen
Gesellschaft} {\bf 18}, p. 318; ----- 1916. {\it Mitteilungen der
Physikalischen Gesellschaft zu Z\"{u}rich} {\bf 16}, p. 47.
\par
9.\ This dual character of the eletromagntic radiation has been
just proposed by Stark, in 1909, in the paper published in the
{\it Physikalische Zetischrift} {\bf 10}, p. 902. In this paper
he explained bremsstrahlung.
\par
10.\ DE BROGLIE, L. 1923. {\it Comptes Rendus de l'Academie des
Sciences de Paris} {\bf 177}, pgs. 507; 548; 630; ----- 1924. {\it
Comptes Rendus de l'Academie des Sciences de Paris} {\bf 179}, p. 39;
----- 1925. {\it Annales de Physique} {\bf 3}, p. 22.
\par
11.\ SCHR\"{O}DINGER, E. 1926. {\it Annales de Physique Leipzig} {\bf
79}, pgs. 361; 489; 734; 747.
\par
12.\ BORN, M. 1926. {\it Zeitschrift f\"{u}r Physik} {\bf 37; 38}, pgs.
863; 803.
\par
13.\ See, for instance, the following textes, in which inclusive
can be found the refe\-rences of papers mentioned in the
Introduction:
\par
13.1.\ POWELL, J. L. and CRASEMAN, B. 1961. {\bf Quantum Mechanics}.
Addison Wesley Publishing Company, Incorporation.
\par
13.2.\ HARRIS, L. and LOEB, A. L. 1963. {\bf Introduction to Wave
Mechanics}, McGraw-Hill Book Company, Inc. and Kogakusha Comapny, Ltd.
\par
13.3.\ DAVYDOV, A. S. 1965. {\bf Quantum Mechanics}. Pergamon Press.
\par
13.4.\ DICKE, R. H. and WITTKE, J. P. 1966. {\bf Introduction to Quantum
Mechanics}. Addison Wesley Publishing Company, Incorporation.
\par
13.5.\ NEWING, R. A. and CUNNINGHAM, J. 1967. {\bf Quantum Mechanics},
Oliver and Boyd Ltd.
\par
13.6.\ SCHIFT, L. I. 1970. {\bf Quantum Mechanics}. McGraw-Hill Book
Company, Incorporation.
\par
13.7.\ MERZBACHER, E. 1976. {\bf Quantum Mechanics}. John Wiley and Sons,
Incorporation.
\par
13.8.\ MOURA, O. 1984. {\bf Mec\^anica Qu\^antica}. EDUFPA.
\par
13.9.\ SHANKAR, R. 1994. {\bf Principles of Quantum Mecha\-nics},
Plenum Press.
\par
14.\ See textes cited in the Note (13).
\par
15.\ BUTKOV, E. 1973. {\bf Mathematical Physics}, Addison-Wesley
Publishing Company.
\par
16.\ FEYNMAN, R. P. and HIBBS, A. R. 1965. {\bf Quantum Mechanics and
Path Integrals}, McGraw-Hill Book Company.
\par
17.\ SILVA, D. G. da 2006. {\bf C\'alculo dos Invariantes de
Ermakov-Lewis e do Pacote de Onda Qu\^antico da Equa\c{c}\~ao de
Schr\"{o}dinger para Medidas Cont\'{\i}nuas}. {\it Trabalho de
Conclus\~ao de Curso},\ DFUFPA.
\par
18.\ NASSAR, A. B. 1990. {\it Physics Letters A 146}, 89; -----
1999. {\bf Wave Function versus Propagator}. (DFUFPA, mimeo); SOUZA,
J. F. de 1999. {\bf Aproxima\c{c}\~ao de de Broglie-Bohm para
Osciladores Harm\^onicos Dependentes do Tempo}. {\it Tese de Mestrado},
DFUFPA.
\end{document}